\documentclass[12pt]{article}
\begin{document}
\title{\large\bf Ground state of the hard-core Bose gas on lattice I.\\
                 Energy estimates}
\author{Andr\'as S\"ut\H o\\
Research Institute for Solid State Physics \\ Hungarian Academy of
Sciences \\ P. O. Box 49, H-1525 Budapest 114 \\ Hungary }
\date{}
\maketitle
\thispagestyle{empty}
\begin{abstract}

\noindent
We investigate the properties of the ground state of a system of interacting 
bosons on regular lattices with coordination number $k\geq 2$. 
The interaction is a pure, infinite, on-site repulsion. Our concern is
to give an improved upper bound on the ground state energy per site. 
For a density $\rho$ a trivial upper bound is known to be $-k\rho(1-\rho)$.
We obtain a smaller variational bound within a reasonably large
family of trial functions. The estimates make use of a large deviation
principle for the energy of the Ising model on the same lattice.

\end{abstract}

\vspace{1cm}
\vspace{1cm}
\newpage

\section{Introduction}

Bosons on a lattice interacting via an infinite on-site repulsion 
(hard-core bosons) represent a system of double interest.
It is the simplest example of an interacting Bose gas and, thus, 
the most promising candidate for a rigorous treatment of Bose-Einstein 
condensation of interacting particles. 
On the other hand, the model is known to be equivalent to a 
system of ${1\over 2}$ spins \cite{MM} coupled via the $XY$-$\;$ and  
possibly the $Z$-components of neighboring spins and exposed to an external
magnetic field in the $Z$ direction. Ordering of the planar component of
the spins is equivalent to Bose-Einstein condensation or the appearance of
off-diagonal long-range order (ODLRO) in the system of bosons. 
In spite of a long and extensive study the results about ordering are
far from being complete.
Apart from some exceptions, like bounds on the density of the condensate
\cite{Th1} or the discussion of the model on the full graph
\cite{Th2,Pen}, the most interesting and difficult results were formulated
in spin terminology \cite{DLS, KLS}. These works made use of a particular
symmetry, the reflexion positivity. This introduced severe limitations as to
the value of the external field (zero field) and the lattice type (essentially
hypercubic lattices). Translated into the language of the boson gas, ODLRO
was shown at half-filling on hypercubic lattices in the ground state
in and above two dimensions, and for low enough temperatures above two 
dimensions.

In this paper we apply the boson terminology. Let $\bf L$ be an infinite 
lattice which, for the sake of simplicity, will be supposed to be regular
with a constant coordination number (valency) $k$. Throughout the paper
$\Lambda$ denotes a finite part of $\bf L$ equipped with a periodic boundary
condition so as to keep the valency constant (not really essential). The
Hamiltonian we are going to study is
\begin{equation}\label{Ham}
H=-\sum_{\{x,y\}\in E\Lambda}(b^*_x b_y+b^*_y b_x)
\end{equation}
We write $x,y,\ldots$ for the vertices of $\bf L$, and $E\Lambda$ for the 
set of edges of $\Lambda$; $b^*_x$ and $b_x$ create, resp., annihilate a
hard-core boson at $x$. Boson operators at different sites commute with
each other while
\begin{equation}\label{hc}
b_x^*b_x+b_xb_x^*=1
\end{equation}
accounts for the hard-core condition. 
Correspondence with spin models is obtained
by setting $b_x=S_x^-$ and $b_x^*=S_x^+$. The Hamiltonian 
conserves the number of bosons,
\begin{equation}\label{N}
N=\sum_{x\in\Lambda}n_x=\sum_{x\in\Lambda}b_x^*b_x
\end{equation}
and is also invariant under particle-hole transformation. We can, therefore,
fix $N$ so that $\rho=N/|\Lambda|$ is between 0 and ${1\over 2}$. (Here
and below, if $A$
is a finite set, $|A|$ denotes the number of its elements.) 
Our concern in this
paper is to provide nontrivial upper bounds to 
the ground state energy, $E_0$. That such an estimate may be useful in the study
of {\em qualitative} properties of the ground state, was an interesting point 
of \cite{KLS}.
 
Let $X,Y,\ldots$ denote $N$-point
subsets of $\Lambda$, called also configurations.
A convenient basis is formed by the states
\begin{equation}\label{basis}
\phi(X)=\prod_{x\in X}b_x^*\Psi_{\rm vac}
\end{equation}
where $\Psi_{\rm vac}$ is the vacuum state. Variational estimates of
the ground state energy are of the form 
\begin{equation}\label{var}
E_0\leq\langle\psi|H|\psi\rangle/\langle\psi|\psi\rangle\ .
\end{equation}
A trivial choice is
\begin{equation}\label{triv}
\psi=\sum \phi(X)
\end{equation}
where the summation goes over all $N$-point subsets of $\Lambda$. It yields
(cf. Section~2)
\begin{equation}\label{uptriv}
E_0\leq -k|\Lambda|{|\Lambda|-2\choose N-1}{|\Lambda|\choose N}^{-1}
=-k\rho(1-\rho)|\Lambda|+O(1)\ .
\end{equation}
The bound is nothing else than minus the average size of the {\em boundaries}
of the configurations: We call the boundary of $X$ the set of half-filled
edges and denote it by $\partial X$. 
Hence, (\ref{uptriv}) is equivalent to
\begin{equation}\label{avbound}
|E_0|\geq
\overline{|\partial X|}\equiv {|\Lambda|\choose N}^{-1}\sum|\partial X|
\end{equation}
with summation over $N$-point configurations. If $\Lambda$ is a full graph,
i.e., any two sites are neighbors, 
(\ref{triv}) is an exact eigenvector and (\ref{avbound}) holds with equality.
In any other case we have a strict inequality.

The present work is about trying to improve on this bound, i.e., to make
the right-hand side of (\ref{avbound}) larger by an amount {\em proportional
to $|\Lambda|$}. It is important to know that this is possible for any $k$
and any $\rho$. In the opposite case, if $-k\rho(1-\rho)$ were to be
the true ground state energy per site, one could easily conclude that
as in the full graph,
the Hamiltonian (\ref{Ham}) has a product ground state in infinite volume,
\begin{equation}\label{prod}
\Psi=\prod_x (\sqrt{\rho}\ |n_x=1\rangle+\sqrt{1-\rho}\ |n_x=0\rangle) 
\end {equation}
with ODLRO and the value of the order parameter at its theoretical maximum
($\rho(1-\rho)$, cf. \cite{Pen}). 

Since by putting equal
weights on every configuration, as in (\ref{triv}), yields the average boundary
size (\ref{avbound}), we expect that a larger value can be obtained by
giving larger weights to configurations with larger boundaries. Therefore,
we calculate variational bounds by using trial functions of the form
\begin{equation}\label{trial}
\psi_v=\sum v_X\phi(X)\quad\qquad v_X=v(|\partial X|)
\end{equation}
i.e., $v_X$ depending on $|\partial X|$ only, and $v(n)$ being concentrated
on $n>\overline{|\partial X|}$. Computability depends on
our ability to estimate the number of configurations with a given size of
boundary. 
The logarithm of this number is 
the entropy of the Ising model in a microcanonical ensemble with a fixed
magnetization, $\sum_{x\in\Lambda}\sigma_x=|\Lambda|(1-2\rho)$.
Indeed, if we put $\sigma_x=-1$ for $x$ in $X$ and $\sigma_x=1$ elsewhere,
we get an Ising configuration with the corresponding union of contours 
$\partial X$ whose total length, $|\partial X|$, is the 
energy of the Ising configuration. 
In mathematical terms, 
the distribution of $|\partial X|$ satisfies a large deviation principle
whose rate function is, apart from a shift, minus 
the specific entropy of the Ising model, cf. \cite{Pf,DS,LPS}.
The exact form of the entropy is unknown
for two- and higher-dimensional lattices. To circumvent this problem, 
we use an approximate formula for the probability of having a boundary
of length $n$,
\begin{equation}\label{ldp}
P_{\Lambda,N}(|\partial X|=n)\approx 
Z^{-1}\exp\{-{(n-M)^2\over 2D^2}\}\ .
\end{equation}
Here $Z$ is for normalization and
\begin{equation}\label{MD2}
M=\overline{|\partial X|}=k\rho(1-\rho)|\Lambda|\qquad 
D^2=\overline{(|\partial X|-\overline{|\partial X|})^2}\ .
\end{equation}
What is the approximation in the above expression? First, we replace the
smooth and concave specific entropy $s(\epsilon)$, having a maximum at
the specific Ising energy $\epsilon_m=k\rho(1-\rho)$, by a parabola. 
Since the improved bound on $|E_0|/|\Lambda|$ we are going to derive is 
$(1+\delta)\epsilon_m$ with a $\delta$ never exceeding 0.2, cf. Table 4 below,
this seems to be a consistent approximation. Second, we surmise that the second
derivative of $s(\epsilon)$ at the maximum is $-|\Lambda|/D^2$. 
An analogous statement holds true for the large deviations of a sum of
identically distributed independent random variables. Now $|\partial X|$ 
{\em is} the sum of identically distributed random variables, although not 
independent but finitely dependent ones, see Section 3. So this approximation 
is hopefully good. In any case, we use the formula (\ref{ldp}) without looking
for further justification. It may happen that energies of trial states 
depending on $|\partial X|$ only can be quite close to $E_0$. Then
the error of the approximation (\ref{ldp}) may invalidate our estimates as
rigorous upper bounds. They still can be useful as approximate formulas
showing the $k$- and $\rho$-dependence of the ground state energy. We apply
the term `bound' with this reservation.

In Section~2 we present the general setup for the estimates.
In Section~3 we derive an expression for $D^2$ which
seems to be valid for any vertex-transitive graph (whose all vertices
are equivalent). This formula may also be of interest in site percolation
problems and in the approximation
of the entropy of the Ising model in nonvanishing magnetic fields.
In Section~4 we compute variational bounds by using
for $v(n)$ step-, exponential and Gaussian functions. The last two will
be seen to yield improved bounds for any value of $k$ and $\rho$. 
Section~5 summarizes the results and indicates the way they can be extended
to the grand-canonical ensemble.

In the true ground state $v_X$ is not a function of $|\partial X|$ only.
The detailed dependence on $X$ may not be relevant for computing the
ground state energy but is absolutely crucial for qualitative
properties, like ordering.
Wave functions of the kind (\ref{trial}) trivially show an off-diagonal
long-range order. In the true ground state
fluctuations around a function of $|\partial X|$ destroy ODLRO in one
dimension and may decrease the order parameter in higher dimensions. 
A discussion of the ground state wave function will be given in a 
subsequent paper \cite{Sut}.

\vspace{2mm}
This work was supported by the Hungarian Scientific Research Fund (OTKA) 
under Grant No. T 030543.

\section{Setup for the energy estimates}

It seems advantageous to consider the eigenvalue problem of (\ref{Ham}) 
as a problem
of graph theory. The $N$-point configurations are vertices of a huge
graph that we call the power graph of $\Lambda$ of order $N$ and denote by
$G=G_{\Lambda,N}$. Two configurations $X$ and $Y$ form an edge of $G$ if
$Y$ can be obtained from $X$ by moving a single particle from a site of 
$X$ to a neighboring site, unoccupied in $X$. Thus, $X$ and $Y$ have
$N-1$ common sites and differ on an edge of $\Lambda$. If 
$VG$ and $EG$ denote, respectively, the set of vertices and edges of $G$ 
then
\begin{equation}\label{VGEG}
|VG|={|\Lambda|\choose N}\qquad 
|EG|=|E\Lambda|{|\Lambda|-2\choose N-1}={1\over 2}k|\Lambda|
{|\Lambda|-2\choose N-1}\ .
\end{equation}
The boundary $\partial X$ of $X$ can be given a new interpretation as the 
set of neighbors of $X$ in $G$. So if $X$ and $Y$ form an edge then
$Y\in\partial X$ and $X\in\partial Y$. 
The action of $-H$ on $G$ is that of the usual lattice Laplacian with the
exception that there is no subtracted diagonal term. The matrix of $-H$ in the
basis (\ref{basis}) is the adjacency matrix $A=A(G)$ of $G$, that is,
$A_{XY}=1$ if $X$ and $Y$ are neighbors and zero otherwise. 
We are interested in the largest eigenvalue of $A$, $\lambda_1=|E_0|$
and in the corresponding eigenvector, $a=(a_X)$. $G$ is connected
if $\Lambda$ is connected (that we suppose), therefore $A$ is irreducible
(ergodic) and the Perron-Frobenius theorem applies: $\lambda_1$ is 
nondegenerate and largest to absolute value, and $a_X>0$ for all $X$.
We note that $G$ is bipartite if and only if $\Lambda$ is bipartite, and
hence $-\lambda_1$ is an eigenvalue of $A$ if and only if $\Lambda$ is
bipartite. 

We call $n_{\rm min}$, resp., $n_{\rm max}$ the minimum, resp., maximum of 
$|\partial X|$ among the $N$-point configurations. Clearly, 
$n_{\rm min}=O(N^{(d-1)/d})$ if $\bf L$ is a d-dim\-en\-sional lattice, and
($\rho\leq {1\over 2}$)
\begin{equation}\label{nmax}
n_{\rm max}\leq kN=k\rho|\Lambda|\ .
\end{equation}
We have the trivial inequalities
\begin{equation}\label{trivineq}
\overline{|\partial X|}\leq\lambda_1\leq n_{\rm max}\ .
\end{equation}
The first is the variational bound (\ref{avbound}): Setting
$v_X\equiv 1$ we find
\begin{equation}\label{triv2}
{|\langle\psi_v|H|\psi_v\rangle|\over\langle\psi_v|\psi_v\rangle}
={\sum_X\sum_{Y\in\partial X} 1\over\sum_X 1}=\overline{|\partial X|}
\end{equation}
which is the same as
\begin{equation}\label{triv3}
{(v,Av)\over (v,v)}={1\over|VG|}\sum_{X,Y:\{X,Y\}\in EG}1={2|EG|\over|VG|}\ .
\end{equation} 
The upper bound in (\ref{trivineq}) follows from the eigenvalue
equation: Let $X_0$ be a configuration on which $a_X$ reaches its maximum,
then
\begin{equation}\label{trivup}
\lambda_1 a_{X_0}=\sum_{Y\in\partial X_0}a_Y\leq |\partial X_0|a_{X_0}\ .
\end{equation}
We note that for $\rho={1\over 2}$ there is a better upper bound,
$\lambda_1\leq{1\over 4}(k+1)|\Lambda|$ (\cite{DLS}, Theorem C.1). 

In Section~4 we shall see that $n_{\rm max}$ plays a role in optimizing
the lower bound to $\lambda_1$. It is therefore important to know $n_{\rm max}$
exactly. In (\ref{nmax}) there is equality if $\rho$ is small enough.
In particular, 
$n_{\rm max}=k\rho |\Lambda|$ for all $\rho\leq {1\over 2}$
on bipartite lattices, and it is an easy graphical exercise to see that
equality holds for $\rho\leq {1\over 3}$ on the triangular ($k=6$) and on the
Kagom\'e ($k=4$) lattices. Moreover, for both lattices $n_{\rm max}$ is constant 
between the
densities ${1\over 3}$ and ${1\over 2}$: $2|\Lambda|$ for the triangular
and ${4\over 3}|\Lambda|$ for the Kagom\'e lattice. This can be seen from
the following argument.
In general, $-n_{\rm max}$ is the ground state energy
of the {\em antiferromagnetic} Ising model under the restriction that the
magnetization is fixed, $\sum_{x\in \Lambda}\sigma_x=(1-2\rho)|\Lambda|$.
However, we do not need to deal with the 
restriction. In both cases the (unrestricted) ground state is known to be
highly degenerate. Among the exponentially large number of ground state
configurations there are nonmagnetized ones, corresponding to $\rho={1\over 2}$,
others with concentration of down-spins $\rho={1\over 3}$, and between these
two limits $\rho$ can vary by steps of ${1\over |\Lambda|}$. The rule is to
flip zero-energy spins one by one. The common 
energy of all these configurations is easy to compute from the fact
that in each triangle there is precisely one unsatisfied bond. This fixes
the value of $n_{\rm max}$ as given above. 

The variational bound (\ref{var}) reads $\lambda_1\geq B(v)$ where
\begin{equation}\label{Bv0}
B(v)\equiv {(v,Av)\over(v,v)}
=\sum_X v_X\sum_{Y\in\partial X}v_Y/\sum_X v_X^2\ .
\end{equation}
In the estimations below a crucial role is played by the
inequality
\begin{equation}\label{deltaxy}
||\partial X|-|\partial Y||\leq 2(k-1)\quad\mbox{if}\quad
Y\in\partial X\ .
\end{equation}
When passing from $X$ to $Y$ a neighboring particle-hole pair is interchanged.
For both the particle and the hole 
the number of neighbors of the opposite kind can change by
at most $k-1$, whence (\ref{deltaxy}) follows.
If $v_X=v(|\partial X|)$ and $v(n)$ is a nondecreasing sequence then
\begin{equation}\label{bound1}
B(v)\geq {\sum_{n=n_{\rm min}}^{n_{\rm max}}
n|\Omega_n\!|\ v(n-2k+2)v(n)\over
\sum_{n=n_{\rm min}}^{n_{\rm max}}|\Omega_n\!|\ v(n)^2}
\end{equation}
where $\Omega_n$ denotes the set
of configurations with boundary length $n$. 

Above it is understood that $v(n)=v(n_{\rm min})$ if $n<n_{\rm min}$.
We have computed bounds given by the right member of (\ref{bound1}), using
step-functions and functions with an exponential or a faster increase. In all
cases we have found no improvement with respect to the trivial bound if
$\rho$ was in a neighborhood of $1\over 2$. Apparently, we have lost too much in
the inequality (\ref{bound1}). 

There is, however, a way to compute $B(v)$ in leading order ($|\Lambda|$)
by making a further hypothesis on some details of the 
large deviation principle (\ref{ldp}). Let $\nu_i(X)$
be the number 
of those neighbors of $X$ having a boundary length $|\partial X|+i$.
(So $\sum_i \nu_i(X)=|\partial X|$.) For $v_X=v(|\partial X|)$ we have
\begin{equation}\label{Bv}
B(v)={\sum_n v(n)\sum_{i=-2k+2}^{2k-2}v(n+i)\sum_{X\in\Omega_n}\nu_i(X)
\over\sum_n v(n)^2 |\Omega_n|}\ .
\end{equation}
Equation (\ref{ldp}) is equivalent to 
\begin{equation}\label{ldp1}
|\Omega_n|/|\Omega_m|\approx\exp\{-[(n-M)^2-(m-M)^2]/2D^2\}\ .
\end{equation}
An analogous statement for the ratio of $\nu_i(X)$ and $\nu_j(X)$ 
is certainly wrong for all $X$ separately but may be correct for the ratio of
their sums over $\Omega_n$. In this hope we formulate the hypothesis that
\begin{equation}\label{ldp2}
{\sum_{X\in\Omega_n} \nu_i(X)\over\sum_{X\in\Omega_n} \nu_j(X)}\approx
{e^{-{(n+i-M)^2\over 2D^2}}\over e^{-{(n+j-M)^2\over 2D^2}}}
=e^{-{n-M\over D^2}(i-j)}\quad -2k+2\leq i,j\leq 2k-2\ .
\end{equation}
The third member of (\ref{ldp2}) is obtained by dropping a term of 
order $|\Lambda|^{-1}$. 
With (\ref{ldp}) or (\ref{ldp1}) and (\ref{ldp2}) Eq.~(\ref{Bv}) reads
\begin{eqnarray}\label{Bv2}
B(v)&=&\left(\sum_n e^{-{(n-M)^2\over 2D^2}}v(n)^2\right)^{-1}
       \sum_n n e^{-{(n-M)^2\over 2D^2}}v(n)\langle v\rangle_n
\\ 
\langle v\rangle_n&=& 
\left(\sum_{j=-2k+2}^{2k-2} e^{{n-M\over D^2}j}\right)^{-1}
\sum_{i=-2k+2}^{2k-2} e^{{n-M\over D^2}i}v(n-i)\ .
\end{eqnarray}
This is the starting point of the estimates of Section~4.

\section{Statistics of the boundary lengths}

In this section we show that the mean square deviation of $|\partial X|$
is given by 
\begin{equation}\label{D2}
D^2=[k-2(2k-1)\rho(1-\rho)]k\rho(1-\rho)|\Lambda|=
[k-2(2k-1)\rho(1-\rho)]M\ .
\end{equation}
We have found this expression first for $d$-dimensional hypercubic lattices,
and checked it later to hold also for the triangular, honeycomb and Kagom\'e 
lattices. This is somewhat surprising because
our derivation below needs knowledge of the rather different local neighborhoods 
up to next-nearest neighbors. The only common feature of all these lattices
seems to be that all sites are symmetry-related and thus equivalent. 
Therefore, it should be possible 
to prove Eq.~(\ref{D2}) on the basis of vertex-transitivity alone.

Equation (\ref{D2}) is obtained by using the grand-canonical probabilities
$P_{\Lambda,\rho}$, i.e., by filling the sites of $\Lambda$ independently
and with equal probability $\rho$. We expect smaller order corrections to appear
if the canonical distribution is used. 
So in this section $X$ is a random subset of $\Lambda$ whose probability
to be selected is $\rho^{|X|}(1-\rho)^{|\Lambda|-|X|}$,
and $n_x=n_x(X)$ is a random variable taking the value 1 if $x$ is in $X$
and 0 otherwise. Then all $n_x$ are independent
and take 1 with probability $\rho$ and 0 with probability $1-\rho$. We define
\begin{equation}\label{fx}
f_x=n_x\sum_{y\in\partial x}(1-n_y)
\end{equation}
where $\partial x$ denotes the set of neighbors of $x$ in $\Lambda$. Clearly,
\begin{equation}\label{fxav}
\overline{f_x}=k\rho(1-\rho)\ .
\end{equation}
The boundary length of $X$ is obtained as
\begin{equation}\label{DeltaX}
|\partial X|=\sum_{x\in \Lambda}f_x(X)\ .
\end{equation}
Thus the mean value of (\ref{DeltaX}) is 
\begin{equation}\label{M}
M=\sum_{x\in\Lambda}\overline{f_x}=k\rho(1-\rho)|\Lambda|
\end{equation}
as found earlier. 

Let $d(x,y)$ denote the graph distance of $x$ and $y$ in $\Lambda$, i.e.,
the length of the shortest walk between them. Since $f_x$ and $f_y$ are
independent if $d(x,y)>2$, we find
\begin{equation}\label{mean}
D^2=\sum_{x,y}\overline{(f_x-\overline{f_x})(f_y-\overline{f_y})}=
\sum_{x,y:\,d(x,y)\leq 2}r(x,y)
\end{equation}
\begin{equation}\label{cov}
r(x,y)=\overline{f_xf_y}-\overline{f_x}^2\ .
\end {equation}

The computation of the different terms is straightforward by observing
that $n_x^2=n_x$, $(1-n_x)^2=1-n_x$ and $n_x(1-n_x)=0$. 
The contribution of the diagonal terms $x=y$ is the same for any $k$-regular
lattice. Namely,
\begin{equation}\label{d=0}
\overline{f_x^2}=k\rho(1-\rho)+k(k-1)\rho(1-\rho)^2
\end{equation}
\begin{equation}\label{c0}
\sum_{x\in\Lambda}r(x,x)=|\Lambda|r(x,x)=
[k-(2k-1)\rho+k\rho^2]M\ .
\end{equation}
The contribution of nearest neighbor pairs depends on the number of triangles
containing a given edge. If there are $\ell$ such triangles then
\begin{equation}\label{d=1}
\overline{f_xf_y}=\rho^2[\ell(1-\rho)+[(k-1)^2-\ell\,](1-\rho)^2]
\end{equation}
\begin{equation}\label{c1}
\sum_{x,y:\,d(x,y)=1}r(x,y)=k|\Lambda|r(x,y)
=\rho[\ell\rho-(2k-1)(1-\rho)]M\ .
\end{equation}
If $x$ and $y$ are next-nearest neighbors to each other, 
they may have $m$ common nearest neighbors. Then
\begin{eqnarray}\label{d=2}
\overline{f_xf_y}&=&\rho^2[m(1-\rho)+(k^2-m)(1-\rho)^2]\nonumber\\
r(x,y)&=&m\rho^3(1-\rho)\ .
\end{eqnarray}
In $d$-dimensional hypercubic lattices ($k=2d$) there are next-nearest
neighbor pairs with $m=1$ and $m=2$. Their contribution to $D^2$ is
\begin{eqnarray}\label{c2Zd}
\sum_{x,y:\,d(x,y)=2}r(x,y)&=&k|\Lambda|\,r(x,y)_{m=1}+4{d\choose2}|\Lambda|\,
r(x,y)_{m=2}\nonumber\\
&=&\rho^2 M+(k-2)\rho^2 M=(k-1)\rho^2 M\ .
\end{eqnarray}
For the triangular lattice ($k=6$)
\begin{equation}\label{c2tr}
\sum_{x,y:\,d(x,y)=2}r(x,y)=k|\Lambda| [r(x,y)_{m=1}+r(x,y)_{m=2}]
=(k-3)\rho^2 M\ .
\end{equation}
In the honeycomb lattice ($k=3$) each site has 6 next-nearest neighbors, 
all of the type $m=1$. So
\begin{equation}\label{c2hon}
\sum_{x,y:\,d(x,y)=2}r(x,y)=2k|\Lambda|\,r(x,y)_{m=1}=(k-1)\rho^2 M\ .
\end{equation}
In the Kagom\'e lattice ($k=4$) there are 8 next-nearest neighbors with
$m=1$:
\begin{equation}\label{c2Kag}
\sum_{x,y:\,d(x,y)=2}r(x,y)=2k|\Lambda|\,r(x,y)_{m=1}=(k-2)\rho^2 M\ .
\end{equation}

Finally, we obtain $D^2$ by adding (\ref{c0}), (\ref{c1}) with $\ell=0$
and (\ref{c2Zd}) for hypercubic lattices,
(\ref{c0}), (\ref{c1}) with $\ell=2$ and (\ref{c2tr}) 
for the triangular lattice, (\ref{c0}), (\ref{c1}) with $\ell=0$ and
(\ref{c2hon}) for the honeycomb lattice and (\ref{c0}), (\ref{c1}) with
$\ell=1$ and (\ref{c2Kag}) for the Kagom\'e lattice. All yield (\ref{D2}).

\section{Energy estimates}

By inspecting Eq.~(\ref{Bv2}) it is clear that $v(n)$ has to be chosen
in such a way that it shifts the expectation value of the Gaussian upwards. The 
appropriate choice can be either a rapidly -- at least exponentially --
increasing function or a function concentrated on values well above
$M$. If in the numerator we had $v(n)$ also at the place of 
$\langle v\rangle_n$, $B(v)=n_{\rm max}$ could be reached. 
However, typically $\langle v\rangle_n/v(n)<1$ and decreases as we modify
$v$ by putting increasing weights to larger boundary lengths.
This limits the maximum of $B(v)$. We repeat here the DLS-bound \cite{DLS},
mentioned in Section 2,
\begin{equation}\label{DLS}
\lambda_1\leq M\left(1+{1\over k}\right)\qquad (\rho={1\over 2})
\end{equation}
and the trivial upper bound (\ref{trivineq}),
\begin{equation}\label{trivineq2}
\lambda_1\leq M{1\over 1-\rho}\ .
\end {equation}
The right members of these inequalities are upper bounds to $B(v)$ as well.

\vspace{2mm}
{\it Step functions}

Let $v(n)=1$ if $n_1\leq n\leq n_2$ and 0 otherwise. Suppose that
$n_1-M=m>c_1|\Lambda|$ and $n_2-n_1>c_2\sqrt{|\Lambda|}$ where 
$c_1$ and $c_2$ are positive constants. We have
\begin{equation}\label{step1}
B(v)=\left(\sum_{n=n_1}^{n_2}e^{-{(n-M)^2\over 2D^2}}\right)^{-1}
\sum_{n=n_1}^{n_2}e^{-{(n-M)^2\over 2D^2}}
\left[n{\sum_ie^{{n-M\over D^2}i}v(n-i)\over 
\sum_ie^{{n-M\over D^2}i}}\right]\ .
\end{equation}
Because of the lower cutoff, the Gaussian is sharply concentrated on $n_1$.
Except for $v(n-i)$, we can therefore replace $n$ by $n_1$ in the square 
bracket. Also, with the above choice of $n_2$, the dependence on $n_2$ is 
negligible. We find after some manipulations that apart from smaller order 
corrections
\begin{equation}\label{step2}
B(v)=n_1\left[1-\left(\sum_{i=-2k+2}^{2k-2}e^{{m\over D^2}i}\right)^{-1}
\sum_{i=1}^{2k-2}e^{{m\over D^2}i}\,
{\sum_{n=n_1}^{n_1+i-1}e^{-{(n-M)^2\over 2D^2}}\over
\sum_{n\geq n_1}e^{-{(n-M)^2\over 2D^2}}}\right] \ .
\end{equation}
We simplify the fraction with $e^{-m^2/2D^2}$ and evaluate
it to find $1-e^{-mi/D^2}$ plus a correction which vanishes as 
$|\Lambda|\rightarrow\infty$. Inserting this into Eq.~(\ref{step2}) we obtain
\begin{eqnarray}\label{step3}
B(v)&=&M\left(1+{m\over M}\right){2k-2+\sum_{i=0}^{2k-2}e^{-{m\over D^2}i}\over
\sum_{i=-2k+2}^{2k-2}e^{{m\over D^2}i}}\nonumber\\
&=&M\left\{1+\left[{D^2\over M}-{(k-1)(2k-1)\over 4k-3}\right]
{m\over D^2}+O\left(\left({m\over D^2}\right)^2\right)\right\}\ .
\end{eqnarray}
From Eq.~(\ref{D2}) we see that $D^2/M$ tends to $k$ as $\rho$ goes to
zero. Hence, the coefficient of $m/D^2$ is positive if $\rho$ is small
enough and we have the best (improved) bound for $n_1$ close to $n_{\rm max}$. 
Then $m/D^2\approx \rho/k$ and we are not in conflict with (\ref{trivineq2}).    
On the other hand, as $\rho$ goes to ${1\over 2}$, $D^2/M$ tends to
${1\over 2}$ and the coefficient of $m/D^2$ in (\ref{step3}) becomes
negative. Thus, for any $k\geq 2$ there is
a neighborhood of $\rho={1\over 2}$ in which the maximum of $B(v)$ is reached
for $m=0$, that is, we get no improved bound.

\vspace{2mm}
{\it Exponential trial functions}

Let $v(n)=e^{xn}$ with an $x>0$. Suppose first $x<x_{\rm max}$ where
\begin{equation}\label{xmax}
x_{\rm max}={n_{\rm max}-M\over 2D^2}\ .
\end{equation}
Plugging $v$ into Eq.~(\ref{Bv2}), due to
the factor $e^{2xn}$ the expectation value of the
Gaussian is shifted from $M$ to $M+2xD^2<n_{\rm max}$ both in the numerator 
and in the denominator. The new expectation value can replace $n$ in its other
occurrences in the numerator. Apart from smaller order corrections we get
\begin{equation}\label{expo1}
B(v)=M\left(1+2xD^2/M\right)F_k(x)\equiv M G_{k,\rho}(x)
\quad (x<x_{\rm max})\ ,
\end{equation}
\begin{equation}\label{Fkx}
F_k(x)={\sum_{i=-2k+2}^{2k-2}e^{xi}\over\sum_{i=-2k+2}^{2k-2}e^{2xi}}
={\cosh(2k-{1\over 2})x-\cosh(2k-{5\over 2})x\over
  \cosh(4k-{5\over 2})x-\cosh(4k-{7\over 2})x}\ .
\end{equation}
Expanding $F_k(x)$ around 0 we find
\begin{equation}\label{Gkrho}
G_{k,\rho}(x)=\left(1+2xD^2/M\right)[1-(k-1)(2k-1)x^2+O(x^4)]>1
\end{equation}
if $x$ is small enough, so we have an improved bound for any $k$ and $\rho$.

If $x>x_{\rm max}$, the maximum of the shifted Gaussian is outside
the range of summation. Therefore the distribution is sharply concentrated
on a small neighborhood of $n_{\rm max}$, and the computation of $B(v)$
can be done in analogy to the case of the step function. The result is
\begin{equation}\label{expo2}
B(v)={n_{\rm max}\over\sum_{i=-2k+2}^{2k-2}e^{2x_{\rm max}i}}
\left[\sum_{i=0}^{2k-2}e^{(2x_{\rm max}-x)i}
+\sum_{i=1}^{2k-2}e^{-xi}\right]
\quad (x>x_{\rm max})\ .
\end{equation}
Notice that $B(v)$ varies continuously with $x$ and both (\ref{expo1}) 
and (\ref{expo2}) yield
\begin{equation}\label{expo3}
B(v)=n_{\rm max}F_k\left(x_{\rm max}\right)
\quad {\rm if}\ x=x_{\rm max}\ .
\end{equation}

In order to
obtain the best bound, $B(v)$ has to be maximized with respect to $x$. 
$G_{k,\rho}(x)$ has a unique maximum at some positive $x$. This can be
the location of the maximum of $B(v)$ only when it is smaller than
$x_{\rm max}$, and then the optimal $B(v)$ is 
computed with it from (\ref{expo1}). Table 1
shows that this is the case for all $k$ if $\rho$ is not too close to 0. 
However, for small enough densities
the maximum of $G_{k,\rho}$ is attained at an $x$ above 
$x_{\rm max}$. For these densities 
the first expression (\ref{expo1}) increases with $x$ up to $x_{\rm max}$
while the second expression (\ref{expo2}) decreases with $x$ for
all densities. So the highest bound is provided by (\ref{expo3}).

The conclusion is that for all densities we can obtain the best bound by
maximizing (\ref{expo1}) with respect to $x$ under the condition that
$x\leq x_{\rm max}$. 

Numerical results on the best estimates are summarized in the tables below. 
They apply to bipartite lattices whenever $D^2$ is given by Eq.~(\ref{D2}).
According to Table 1, they are valid also to the Kagom\'e and triangular 
lattices, because the largest density for which the maximizing $x$ is 
$x_{\rm max}$ remains below ${1\over 3}$ for every $k$.

\vspace{10mm}
\begin{displaymath}
\begin{tabular}{|c||c|c|c|c|c|}\hline
\multicolumn{1}{|c||}{\ } & \multicolumn{5}{c|}{\rm k} \\ \cline{2-6}
$\rho$ & 2 & 3 & 4 & 6 & 8
\\ \hline
0.1 & 0.038$^*$ & 0.026$^*$ & 0.020$^*$ & 0.014$^*$ & 0.010$^*$ \\
0.2 & 0.120$^*$ & 0.089$^*$ & 0.071$^*$ & 0.040 & 0.027 \\
0.3 & 0.206 & 0.082 & 0.047 & 0.024 & 0.016 \\ 
0.4 & 0.167 & 0.057 & 0.030 & 0.013 & 0.008 \\
0.5 & 0.152 & 0.048 & 0.023 & 0.009 & 0.005 \\ \hline
\end{tabular}
\end{displaymath}
{\em Table 1.} Rates of increase, $x$, of the exponential trial functions, 
maximizing $G_{k,\rho}$. Numbers with an asterix correspond to $x_{\rm max}$.

\vspace{10mm}
\begin{displaymath}
\begin{tabular}{|c||c|c|c|c|c|}\hline
\multicolumn{1}{|c||}{\ } & \multicolumn{5}{c|}{\rm k} \\ \cline{2-6}
$\rho$ & 2 & 3 & 4 & 6 & 8
\\ \hline
0.1 & 1.106 & 1.103 & 1.102 & 1.099 & 1.098 \\
0.2 & 1.199 & 1.157 & 1.128 & 1.100 & 1.088 \\
0.3 & 1.155 & 1.075 & 1.051 & 1.033 & 1.027 \\
0.4 & 1.094 & 1.035 & 1.019 & 1.009 & 1.006 \\
0.5 & 1.077 & 1.024 & 1.012 & 1.004 & 1.002 \\ \hline
\end{tabular}
\end{displaymath}
{\em Table 2.} Maximal values of $G_{k,\rho}(x)$, computed with the 
entries of Table 1. 

\vspace{10mm}
\begin{displaymath}
\begin{tabular}{|c||c|c|c|c|c|}\hline
\multicolumn{1}{|c||}{\ } & \multicolumn{5}{c|}{\rm k} \\ \cline{2-6}
$\rho$ & 2 & 3 & 4 & 6 & 8
\\ \hline
0.1 & 0.199 & 0.298 & 0.397 & 0.594 & 0.791 \\
0.2 & 0.383 & 0.555 & 0.722 & 1.056 & 1.393 \\
0.3 & 0.485 & 0.677 & 0.882 & 1.302 & 1.725 \\
0.4 & 0.525 & 0.745 & 0.978 & 1.453 & 1.932 \\
0.5 & 0.538 & 0.768 & 1.012 & 1.507 & 2.005 \\ \hline
\end{tabular}
\end{displaymath}
{\em Table 3.} Best estimates of $|E_0|/|\Lambda|$ obtained by multiplying
$k\rho(1-\rho)$ with the corresponding entry of Table 2.

\vspace{10mm}
\begin{displaymath}
\begin{tabular}{|c||c|c|c|c|c|} \hline
k&2&3&4&6&8 \\ \hline
$\rho$ & 0.23 & 0.19 & 0.17 & 0.16 & 0.15 \\
$x$ & 0.158 & 0.080 & 0.051 & 0.031 & 0.021 \\
$G_{k,\rho}(x)$ & 1.206 & 1.159 & 1.143 & 1.129 & 1.123 \\ \hline 
\end{tabular}
\end{displaymath}
{\em Table 4.} The maximum of $G_{k,\rho}(x)$ as a function of 
$\rho$ and $x\in[0,x_{\rm max}]$, together with the maximizing $\rho$ and
$x$. In each case the latter is $x_{\rm max}$.

\vspace{10mm}
{\it Gaussian trial functions}

Let $v(n)=e^{-(n-M_1)^2/4D_1^2}$ where $M_1>M$ and $D_1^2\rightarrow\infty$ 
with increasing $|\Lambda|$ but otherwise are free parameters.
The product of the two Gaussians gives rise to a new Gaussian with mean value
and variance
\begin{equation}\label{mudel}
M_2={MD_1^2+M_1D^2\over D_1^2+D^2}\qquad D_2^2={D_1^2D^2\over D_1^2+D^2}\ .
\end{equation}
A straightforward computation yields the same $B(v)$, Eqs.~(\ref{expo1}), 
(\ref{expo2}) and (\ref{expo3}) as
for the exponential trial function with
\begin{equation}\label{Gauss}
x={1\over 2}{M_1-M\over D_1^2+D^2}\ .
\end{equation}
So Gaussian and exponential trial functions provide the same bound, at least
in leading order. Moreover, for any fixed $x>0$ there is a one-parameter 
family of Gaussians giving the same result. An interesting choice is
$D_1^2=o(|\Lambda|)$, e.g. $D_1^2={\rm const}\sqrt{|\Lambda|}$. In this
case $M_2/M_1=1$ and $D_2^2/D_1^2=1$ asymptotically. 

\section{Summary and concluding remarks}

We have presented variational estimates of the ground state energy of a gas
of hard-core bosons on regular lattices. The wave functions we have used
depended only on the size of the boundary of the $N$-point configurations.
Therefore, the estimates could be based on a large deviation principle
governing the distribution of the boundary sizes. The corresponding rate
function is related to the entropy of the (ferromagnetic) Ising model
on the same lattice. We have derived a formula
for the mean-square deviation of the boundary sizes and applied it in a
quadratic approximation of the rate function.

The best estimates we have found are of the form (\ref{expo1}) where
for given $k$ and $\rho$ the parameter $x$ is to be determined numerically 
so as to maximize $B(v)$. This has been illustrated in Tables 1-3.
The maximum is realized either by an exponential
function or by a one-parameter family of Gaussian functions.

Extension to the grand-canonical ensemble is straightforward. Adding 
$-\mu\sum_{x\in\Lambda}n_x$ to the Hamiltonian resumes in adding 
$\mu\rho|\Lambda|$
to $B(v)$. After maximizing with respect to $x$ we have to maximize with
respect to $\rho$. More precisely, if $b(\rho)$ is the maximum of 
$B(v)/|\Lambda|$ with respect to $x$, first we extend $b(\rho)$ symmetrically
to $\rho>{1\over 2}$ and then determine $\hat{b}(\mu)=\max_\rho[b(\rho)+\mu\rho]$
to get an estimate of the ground state energy per site in the
full bosonic Fock space. It is more $\hat{b}(\mu)$ than $b(\rho)$ which is
useful for the equivalent spin model whose Hamiltonian now contains an external
magnetic field in the $Z$-direction.
The function $\rho(\mu)$ that we can find in determining $\hat{b}(\mu)$
is only an approximation of the true relationship which exists between 
the chemical potential and the density in the ground state. Table 3 suggests
that $\rho(0)={1\over 2}$. This rigorously holds true for any positive 
temperature; even we have a much stronger `uniform density theorem',
as in the Hubbard model \cite{LLC}: Because of the particle-hole symmetry,
\begin{equation}\label{unif}
{{\rm Tr}\,n_xe^{-\beta H_{\mu=0}}\over{\rm Tr}\,e^{-\beta H_{\mu=0}}}=
{{\rm Tr}\,(1-n_x)e^{-\beta H_{\mu=0}}\over{\rm Tr}\,e^{-\beta H_{\mu=0}}}=
{1\over 2}
\end{equation}
is valid in any $\Lambda$ with free boundary condition.

In our estimates the lattice structure appears only through $k$, the
coordination number. Therefore we obtain the same result for the Kagom\'e
lattice as for the square lattice, and for the triangular lattice as for
the simple cubic lattice. Although the 
energy corresponds to a nearest neighbor correlation,
the details of the lattice geometry certainly influence the $\rho$-dependence
of the exact ground state energy per site. The mark of the lattice could be 
recovered by using the exact Ising entropies instead of their quadratic
approximants.

\end{document}